\documentstyle{mn}
\title{Intrinsic intraday variability in the gravitational lens system 
B0218+357}
\author[A.~D.~Biggs et al.]{A.~D.~Biggs,\thanks{E-mail: adb@jb.man.ac.uk} 
I.~W.~A. Browne and P.~N.~Wilkinson\\
University of Manchester, Jodrell Bank Observatory, Macclesfield, Cheshire 
SK11 9DL\\}

\begin{document}
\maketitle
\begin{abstract}

Radio monitoring of the gravitational lens system B0218+357 reveals it to 
be a highly variable source with variations on timescales of a few days 
correlated in both images. This shows that the variability is intrinsic to 
the background lensed source and suggests that similar variations in other 
intraday variable sources can also be intrinsic in origin.

\end{abstract}


\section{Introduction}

The origin of the intraday variability (IDV) \cite{wagner95} seen in many 
flat-spectrum radio sources is a major astrophysical puzzle. Such rapid 
variability, if intrinsic to the source, implies a very small source linear 
extent and correspondingly extreme values for the source surface brightness 
temperature, $T_{\mathrm{b}}$. For example, in the BL~Lac object 0716+714 
\cite{wagner96}, where an intrinsic origin is favoured due to the correlation 
between the optical and radio variations, the implied brightness temperature 
is of order $10^{17}$K, well in excess of the inverse Compton limit of 
$\sim$10$^{12}$K for incoherent synchrotron sources. Invoking relativistic 
boosting can reduce the implied brightness temperatures, but often the 
Doppler factors required are so high ($\ga$100) as to be theoretically 
uncomfortable \cite{begelman94} and inconsistent with the statistics of 
superluminal motion \cite{vermeulen94}. 

Extrinsic effects that can cause rapid flux density variations include 
interstellar scintillation (ISS) \cite{rickett90} and gravitational 
microlensing \cite{gopal91}. Whilst the latter has not been suggested as a 
major contributor to the IDV of any source to date, ISS has recently been 
successful in explaining the rapid variability of several sources, perhaps 
most notably the quasars J1819+3845 \cite{dennett00} and PKS~0405-385
\cite{kedziora97}. Where ISS is the dominant cause of the variations the 
source brightness temperature can be reduced, to that requiring a reasonable
Doppler factor ($\sim$10) in the case of J1819+3845. However, as the source 
size required for scintillation is itself very small (of order $\mu$arcsec), 
the source brightness temperature can still greatly exceed the Compton limit, 
by several orders of magnitude in the case of PKS~0405-385.

Even though both intrinsic and extrinsic explanations have their difficulties, 
it would represent considerable progress if it could be proven that the 
variations had one or the other origin. The study of radio sources which have 
multiple gravitational images offers an elegant way to do this. If an IDV 
source is multiply imaged, only intrinsic variations should be correlated in 
the images.

\begin{figure}
\begin{center}
\setlength{\unitlength}{1cm}
\begin{picture}(5,9)
\put(-2.4,-2){\includegraphics{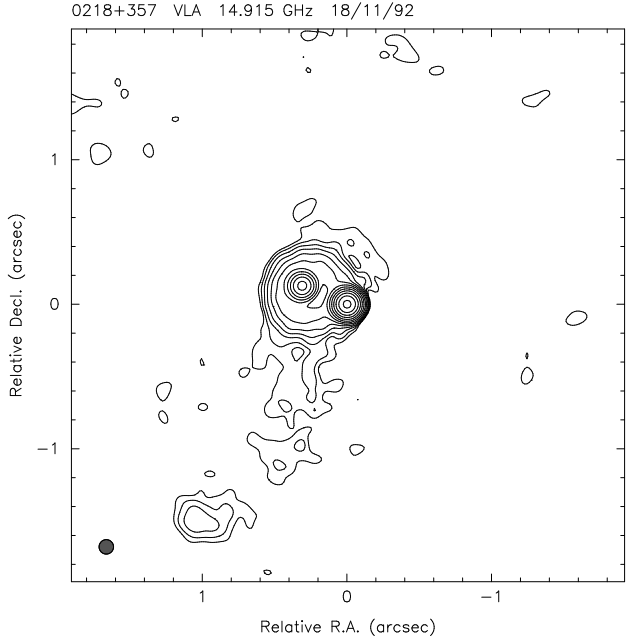}}
\end{picture}
\caption{VLA 15-GHz radio map of B0218+357. As well as the two compact 
components (A to the right) and the Einstein ring, also clearly visible 
is a (non-lensed) radio jet to the south.}
\label{0218map}
\end{center}
\end{figure}

The gravitational lens system B0218+357 \cite{patnaik93} has been subject to
many monitoring campaigns with the ultimate goal of measuring the Hubble 
parameter using the method of Refsdal (1964). The system consists of two 
images (A and B separated by 335~mas) of a compact flat-spectrum radio source 
and a steep-spectrum Einstein ring (Fig.~\ref{0218map}). The redshift of the 
background lensed object is 0.96 \cite{lawrence96}. The flux densities of 
A and B are approximately 800~mJy and 200~mJy respectively and typically have 
$\sim$10~per~cent linear polarizations. We have recently measured a time delay 
of $10.5\pm0.4$~d in this system exploiting its rapid radio variability 
\cite{biggs99}. In this note we use our existing monitoring data and show 
that variability with an implied brightness temperature of $\gg$10$^{12}$K 
is correlated in the two images. We argue that this cannot be due to ISS in
the Galaxy or microlensing in the lensing galaxy and must therefore be 
intrinsic to the background source.

\section{Observations and light curves} 

\begin{figure*}
\begin{center}
\setlength{\unitlength}{1cm}
\begin{picture}(10,6)
\put(5.5,-0.5){\includegraphics{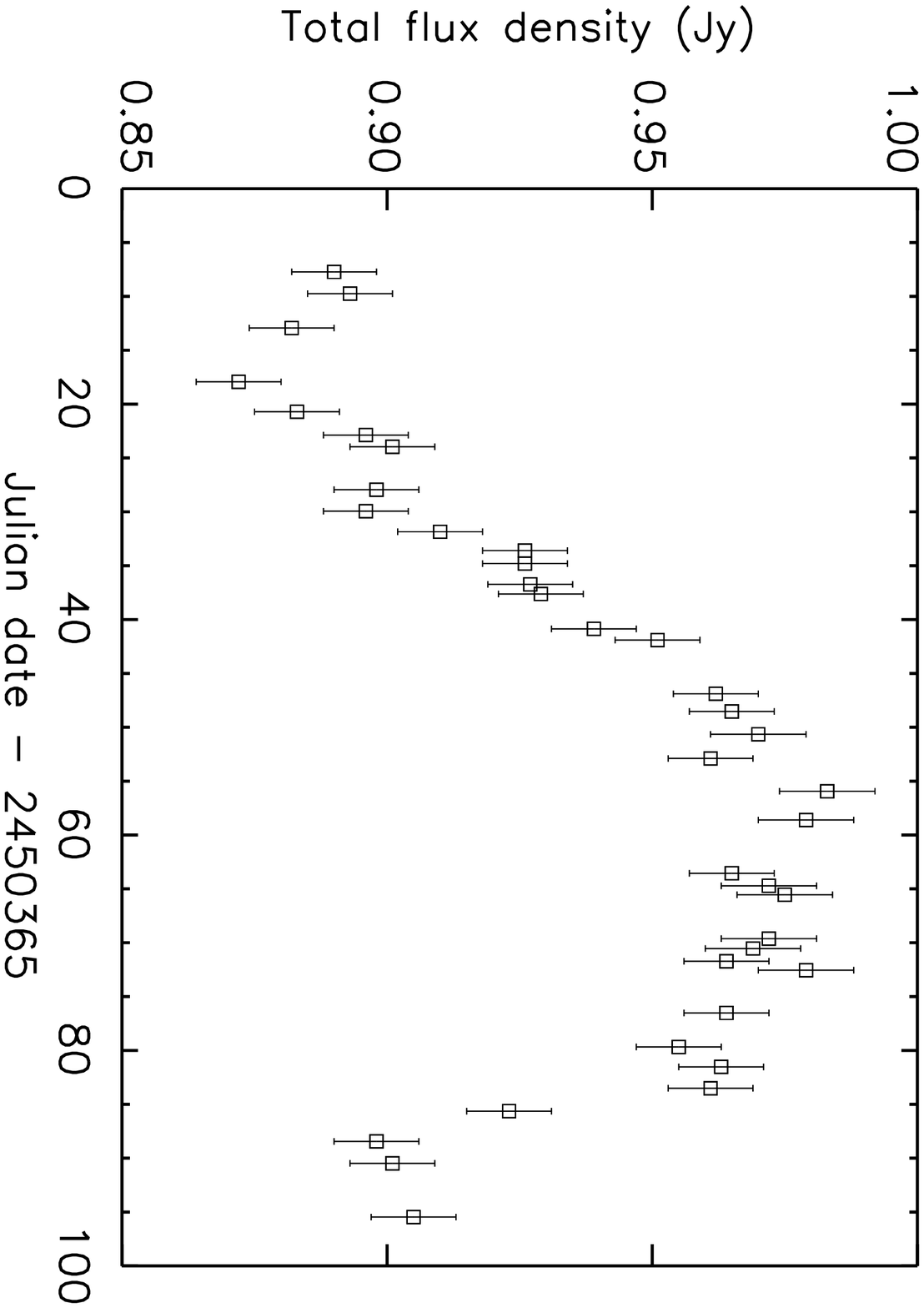}}
\put(14.5,-0.5){\includegraphics{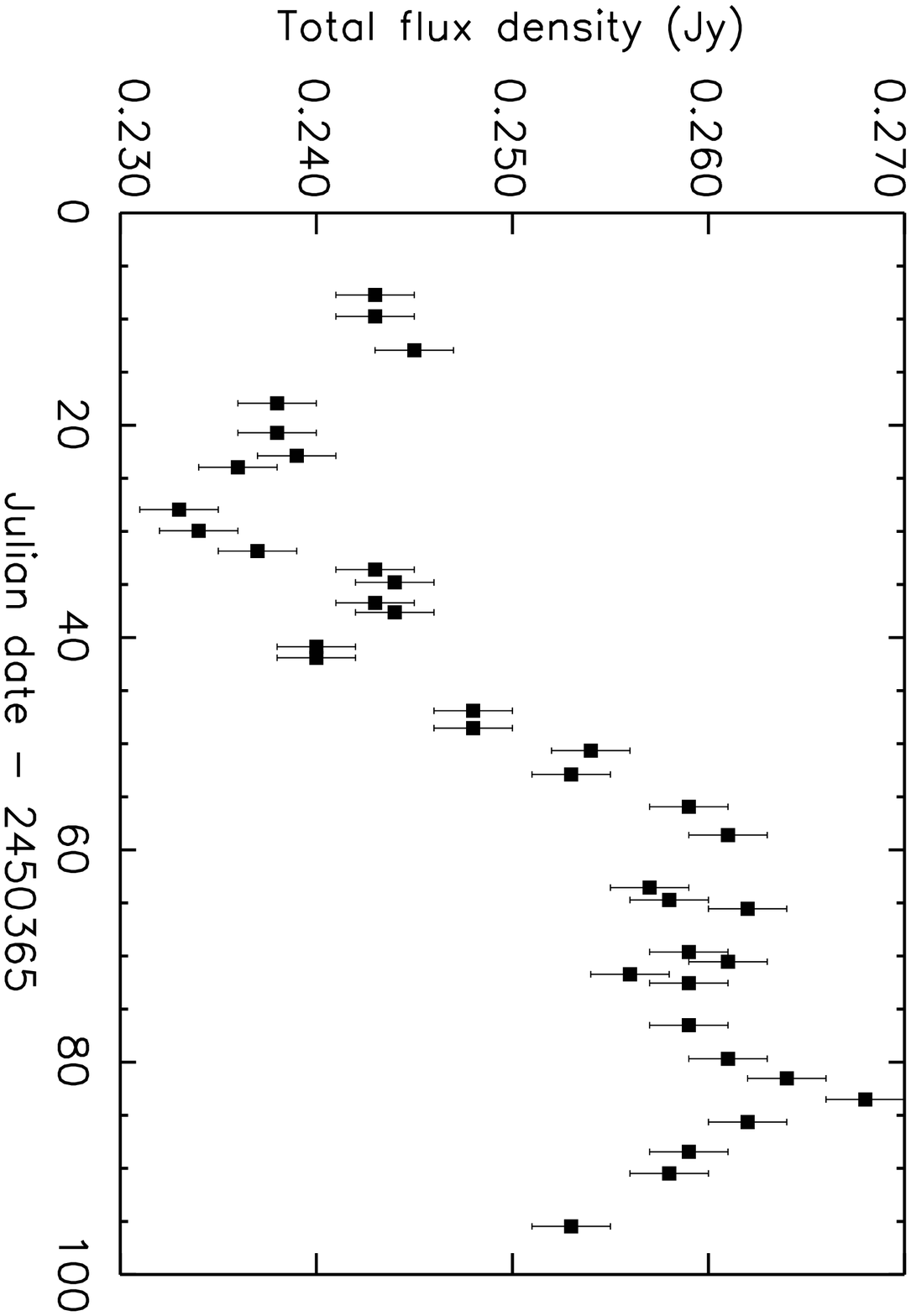}}
\end{picture}
\caption{VLA 15-GHz total flux density light curves. Component A - left, 
component B - right.}
\label{Ugraphs}
\end{center}
\end{figure*}

\begin{figure}
\begin{center}
\setlength{\unitlength}{1cm}
\begin{picture}(10,6)
\put(9,-0.75){\includegraphics{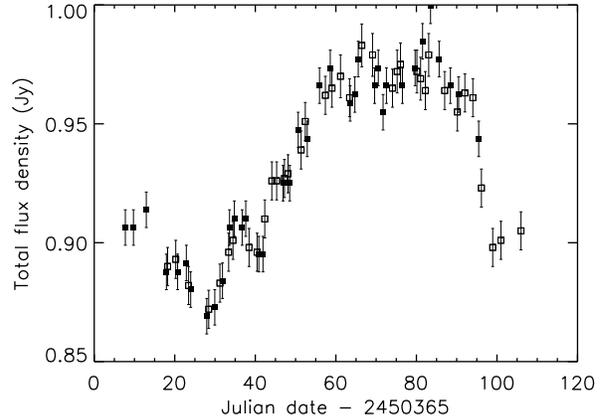}}
\end{picture}
\caption{VLA 15-GHz total flux density light curve constructed from combining
the component A (open squares) and component B (filled squares) data shifted 
by a time delay of 10.5~d and scaled by the flux density ratio (A/B) of 3.73.}
\label{Ucomb}
\end{center}
\end{figure}

B0218+357 was observed with the VLA in `A' configuration between the months 
of 1996 October and 1997 January. Observations were taken at two frequencies, 
15~GHz and 8.4~GHz. In all, data were obtained at 47 epochs, with an average
spacing between observations of $\sim$2~d. Full details of the observations 
and the data reduction are given in Biggs et al. (1999). Although the
light curves are similar at both frequencies, it is at 15~GHz that the
shortest timescale variability is most pronounced. Therefore we
concentrate on the higher-frequency data in this paper and in
Fig.~\ref{Ugraphs} we reproduce the 15-GHz light curves for the two
images. The time delay of $10.5\pm0.4$~d was determined from these and
other light curves using a chi-squared minimisation technique along
with a Monte Carlo analysis of simulated light curves to derive the
associated error. Knowing the delay and the relative magnifications of
the images, it is possible to remove the time delay from the light
curve of one of the images and produce a combined light curve. This is
shown in Fig.~\ref{Ucomb}. It is clear from a visual inspection of the
combined light curve that the variations in the two images correlate
very well. More formally, the time delay analysis of Biggs et
al. (1999) showed that the reduced chi-squared ($\overline{\chi}^2$)
of the fit was 0.7, indicating an excellent fit.

We have no direct evidence for variability from the monitoring data on 
time-scales less than the typical sampling interval of $\sim$2 days. However,
in addition to the VLA monitoring data we have a VLA (also `A' configuration)
15-GHz 10~hr dataset of B0218+357 from which the map in Fig.~\ref{0218map} 
was made. Observations of B0218+357 were taken in scans of approximately 
20~minutes duration, each separated by a $\sim$2~minute scan of the 
phase-calibration source B0234+285. This source was also used to calibrate
the instrumental polarization residuals. 

In order to study the short-term variability properties of B0218+357 we have 
separated the data into the individual scans (of which there are 20) before 
splitting each of these further into four equal sub-scans. This gives a total 
of 80 epochs of data which were then model-fitted in {\sc difmap} 
\cite{shepherd97} and in {\sc aips} using the {\sc uvfit} task. The latter 
was used for fitting to the Stokes $Q$ and $U$ parameters as well as to check 
on the {\sc difmap} Stokes $I$ results. The model-fitting was simplified by 
filtering the data in the $(u,v)$ plane so as to remove the short baselines 
that respond to the diffuse Einstein ring emission. This Fourier 
filtering technique enabled us to restrict the model to two point-like 
components and a good fit ($\overline{\chi}^2 \sim 1$) obtained. 

In order to avoid as much as possible errors in the flux calibration we have 
only considered the flux density {\em ratio} between A and B, the percentage 
polarization of each image and the difference in the polarization position 
angle between the two components. These quantities should all be relatively 
free from systematic calibration errors. The uncertainty in our results has 
been estimated by calculating a mean and a standard deviation from the four 
epochs that make up each scan. We do not show any of the `light curves' 
derived from this process as we do not believe that they show any evidence for
significant variability ($<1$~per~cent) over the 10~hr period. 

\section{Discussion}

Does B0218+357 qualify as an intraday variable radio source? The variability 
in this lens system is less dramatic than that seen in the sources 0716+714 
\cite{wagner96} and 0917+624 \cite{rickett95}. An even greater contrast 
exists between B0218+357 and the quasars PKS~0405-385 \cite{kedziora97} and
J1819+3845 \cite{dennett00} where much greater flux density changes (over
300~per~cent in J1819+3845) occur on timescales of hours. However, the flux 
density of B0218+357 does vary significantly over the typical time-scale of 
our monitoring observations, $\sim$2~d, which is equivalent to $\sim$1~d in 
the rest-frame of the source. In particular we draw attention to 
the rapid fall in the A light curve that occurs towards the end of the 
monitoring where a decrease of about 60~mJy in the total flux density occurs 
over a period of around 5~d. This is associated with a rapid change in 
percentage polarization of about 1~per~cent. Unfortunately, both for the time 
delay measurement and the present discussion, the event is too close to the 
end of the monitoring period for us to be able to see the corresponding 
feature in the B curve. Events that are obviously correlated in the light 
curves of Fig.~\ref{Ugraphs} occur in between days 20 and 40, the biggest of 
which is a change of about 40~mJy over 10~d.

From the observed variations it is possible to estimate minimum brightness
temperatures for the varying part of the background source. For the purposes 
of this discussion we shall restrict ourselves to consideration of the most 
prominent and defined feature in the light curves, the change in flux density, 
$\Delta S$, of 60~mJy near the end of the monitoring. Due to the uncertainty 
in defining the timescale of flux density variations, $t$, we have estimated 
this parameter in several different ways. Firstly, and most simply, we have 
taken $t$ to be equal to the duration of the feature, $\Delta t$, as measured 
directly from the light curves i.e. approximately 5~d. Another approach is to 
weight $\Delta t$ by the flux density of the varying part of the background 
source, $S_{\mathrm{vary}}$ i.e. $t = \Delta t \left|S_{\mathrm{vary}}/\Delta 
S\right|$ (e.g. Wagner \& Witzel 1995). What value should we assign to 
$S_{\mathrm{vary}}$? The conservative approach in terms of brightness 
temperature estimation would be to set this equal to the flux density of the 
component, $\sim1$~Jy in the case of component A. However, VLBI observations 
\cite{patnaik95} have shown that approximately one third of the flux density 
of components A and B originates from a jet-like subcomponent that is unlikely 
to vary significantly over the course of our monitoring observations. 
Therefore, $S_{\mathrm{vary}}$ and $t$ are reduced accordingly.

We have converted our variability timescales into brightness temperatures 
using standard formulae such as equation~(8) of Gopal-Krishna \& Subramanian 
(1991), transforming to a frame comoving with the source. Distance 
measurements were calculated assuming a Hubble parameter of 
65\,km\,s$^{-1}\,$Mpc$^{-1}$ and a flat universe with $\Omega = 1$. A 
correction has also been made for the lens magnification using the model of 
Biggs et al. (1999). Although not stated in that paper explicitly, the 
absolute magnifications of components A and B are equal to 2.1 and 0.6 
respectively (L.V.E.~Koopmans, private communication). We find values of 
$T_{\mathrm{b}}$ in the range $10^{14} - 10^{15}$K. Brightness temperatures 
as high as these are greatly in excess of the inverse Compton limit of 
approximately $10^{12}$K for incoherent synchrotron radiation. This can be 
reconciled by assuming bulk relativistic motion within a radio jet of Doppler 
factor $D$. Variability-derived brightness temperatures are then reduced by a 
factor $D^3$ \cite{readhead94} which for B0218+357 would require a Doppler 
factor of $\leq$10. This is consistent with observations of superluminal 
motion observed in the general source population \cite{vermeulen94}, but 
regular VLBI observations have yet to reliably detect any change in the 
core-jet substructure of the A and B images of B0218+357 (e.g. Porcas \& 
Patnaik 1996).

Could the correlated variability seen in the images of B0218+357 be produced 
by any of the extrinsic mechanisms? First let us consider interstellar 
scintillation. Let us assume that a typical velocity of the local Galactic 
interstellar medium with respect to the Earth is 20~km~s$^{-1}$. To produce 
correlated changes in images A and B with a time lag of $\leq$10~d the screen 
would have to move $\geq$335~mas in this time. With the above velocity the
scintillating screen would have to be at a distance of less than 0.3~pc in 
order to produce correlated variations. This then allows us to
discount the ISS explanation in two ways, the first of which is to note
that 0.3~pc is very much less than the scale height of the Galactic
disk. Secondly, at this distance and at this frequency the angular
size of the first Fresnel zone is 0.1~mas (e.g. Walker 1998). Any
source smaller than this would necessarily scintillate, but this has
not been observed (VSOP calibration sources for example;
J.~E.~J. Lovell, private communication). Walker (1996) has also 
concluded that as the typical separations ($\sim$1~arcsec) of
galactic-scale mass gravitational lens systems are large compared to
the typical size of interstellar irregularities, any scintillation of
the multiple images will be independent. Although the microlensing
explanation might appear tempting, due to the very obvious presence of
a lensing galaxy between us and the background source, the lines of
sight when passing through this galaxy are so far separated that a
correlated time delay of $\sim$10~d can be completely ruled
out. Microlensing in the host galaxy of the background radio source
can be discounted for the same reason as scintillation in the Galaxy.

\section{Conclusions}

The source in the lens system B0218+357 is bright and rapidly variable with
variations on a timescale of a few days that are correlated in both images. 
Thus on these timescales, at least, the mechanism producing the major 
variations must be intrinsic and any contribution from interstellar 
scintillation small. Therefore, although recent results have favoured an 
extrinsic origin for the intraday variations seen in some sources 
(particularly J1819+3845 and PKS~0405-385), our results would 
suggest that IDV in these and other objects can also be partly intrinsic. For 
example, 230~GHz observations of PKS~0405-385 \cite{wagner98} have shown 
rapid variability (with an associated brightness temperature of $10^{14}$K) 
that cannot be due to scintillation. Clearly more intensive monitoring of 
B0218+357, possibly in bursts separated by the time delay, would yield even 
tighter constraints on the intraday variability and enable the time delay to 
be refined. The same argument used here could also be applied to other 
variable lens systems, for example PKS~1830-211 \cite{jauncey91,lovell98}.

\section*{Acknowledgements}

ADB acknowledges the receipt of a PPARC studentship. We thank the
referee, Jim Lovell, for helpful comments. This research was supported
in part by the European Commission TMR Programme, Research Network
Contract ERBFMRXCT96-0034 `CERES'. The VLA is operated by the National
Radio Astronomy Observatory which is a facility of the National
Science Foundation operated under cooperative agreement by Associated
Universities, Inc.

\end{document}